# A Study of the Management of Electronic Medical Records in Fijian Hospitals

**Swaran S. Ravindra[1*] & Rohitash Chandra[2]* & Virallikattur S. Dhenesh[1*]**

1 School of Computing, Information and Mathematical Sciences, University of the South Pacific, Laucala Campus, Fiji

2 Artificial Intelligence and Cybernetics Research Group, Software Foundation, Nausori, Fiji
*Authors are in order of contribution.

Email addresses:

SSR: swaran_ravindra@yahoo.com

RC: c.rohitash@gmail.com

VSD: vsdhenesh@gmail.com

**Key Words:**

Australian Agency for International Development (AusAID) – is an Australian agency that manages development and assistance projects internationally. AusAID has recently been absorbed into the Australian Department of Foreign Affairs and Trade [1] .

Biomedical Informatics – is the field of science that develops theories, techniques, methods pertaining to the use data, information and knowledge which support and improve biomedical research, human health, and the delivery of healthcare services [2] .

Cloud Computing- refers to Information Technology services leased to a person or organization over internet network according to service level requirements. It requires minimal management effort or service provider interaction [3]

e-Health- an emerging field in the intersection of medical informatics, public health and business, referring to health services and information delivered through the Internet and related technologies [4].

Electronic Medical Record (EMR)- An electronic medical record (EMR) is a digital version of a patient's medical history [5].

Health Informatics - The field of information science pertaining to the management of healthcare data and information by the application of computer related technologies [2].

Open Source Software (OSS)- refers to software whose codes are available to be modified as per the users' needs [6]

OpenEMR- is an open source medical practice management application which features a fully integrated EMR system with other features such as scheduling, electronic billing, free support, with the ability to work across different platforms [7].

World Health Organisation (WHO)- is a specialised United Nations agency whose primary role is to direct and coordinate international health within the United Nations' system [8].

## Abstract

Despite having a number of benefits for healthcare settings, the successful implementation of health information systems (HIS) continues to be a challenge in many developing countries. This paper examines the current state of health information systems in government hospitals in Fiji. It also investigates if the general public as well as medical practitioners in Fiji have interest in having web based electronic medical records systems that allow patients to access their medical reports and make online bookings for their appointments. Nausori Health Centre was used as a case study to examine the information systems in a government hospital in Fiji.

## Goals:

i) To study the current state of health information systems and EMR in various Fijian government hospitals, using Nausori Health Centre as a case study

ii) To ascertain if the people (health practitioners and general public) of Fiji are ready for a web based electronic medical records system

## Results

A series of surveys and interviews were conducted to review current systems. The results show that majority of the people surveyed would like to make their appointments and access their medical reports through a web based EMR systems. It also indicated that government hospitals in Fiji are generally lacking computers and medical practitioners and patients would like to have 24/7 access to an integrated web based EMR system.

## Conclusions

Nausori Health Center still has many manual practices that is causing unnecessary delays for both patients and workers. Majority of the staff at Nausori Health Center would agree to a need for a complete EMR system that is accessible anywhere in Fiji which can handle patient records and reporting. Patients and medical practitioners are interested in a web based EMR system for government hospitals that has a patient portal so patients can access their reports and make appointments for their check-up themselves. More studies need to be carried out regarding options for web based EMR systems that would meet the requirements of a developing nation such as Fiji.

## Background

Electronic medical record systems in hospitals have significant benefits [9] [10] [11]. Not only does it reduce the provisions for errors evident in typical manual operations [11], it also promotes an environmentally friendly paperless environment [12] which improves communication of information amongst medical professionals [13]. This allows timely access to patient records for easier decision making in critical situations [14] where the expertise and consultation of distant specialists may be required [15]. Other benefits include significant reduction of errors, elimination of legibility issues [16], easier billing methods and having a data repository for future research and quality improvement [17]. Essentially, an EMR has the ability to facilitate the continuity of care.

EMR systems subsequently assist the organisation and the society as a whole [10] [11]. EMR systems provide many benefits to hospitals and health centers in managing their daily operations. However, just as many other information systems, EMR adaption and implementation prove to be more successful when users are involved in the discussion for its design [18] and implementation. Developing countries such as India, Kenya and Haiti have benefited greatly with the intervention of EMR systems which provides accuracy, efficiency and has overall cost benefits [19].

However, the benefits of an effective EMR system lie heavily in its successful implementation. For EMR implementation to be successful, developing countries need to ensure they are equipped with resources such as adequate infrastructure, finance, staffing, computers, training and computer literacy and careful strategic and tactical planning [20].

It is important to understand that there are some notable disadvantages of EMRs, some of which include investment into technology and its upkeep, transition time, privacy and security issues [21] [11]. Some practitioners find it easier to scribble notes on a paper [22] and update later rather than doing real time data entry as they find it time consuming and feel that it takes away the human factor in communication with patients [23] [24] [25]. However, one study suggests that the time spent on entering data in an EMR can actually give time to the patient to think of any other questions or queries they may have, which could be seen as a positive effect of EMRs [26].

There are some known challenges of implementation of EMRs. Since healthcare facilities are complex and interdependent [27], work flow and methods need to be carefully mapped to the EMR system or else implementation can prove to be a challenge [28]. There is a lack of research in design strategies for EMR systems particularly for developing countries [27]. A general lack of promotion of health information system, lack of resources such as infrastructure, finance, technology, experienced personnel and medical professionals' resistance towards technology are also some challenges that have been noted in developing countries [29].

*Health Informatics,* used synonymously with the term *Biomedical Informatics*, is an emerging field of information science that concerns management of general healthcare data and information by utilising computers and related technologies [2]. Despite having a number of benefits for healthcare, the successful implementation of health information systems (HIS) continues to be a challenge in many developing countries. Some of these challenges include 1) human barriers such as belief systems, behaviours and attitudes, 2) professional barriers related to the nature of healthcare jobs, 3) technical barriers related to the use of computers, 4) organizational barriers such as the hospital management, 5) financial barriers related to funding, 6) and legal and regulatory barriers [29]. Political issues and conflict of interest amongst medical professionals are other barriers particularly in developing countries. The World Health Organization identifies financial constraints and lack of technical expertise as barriers to health information systems in developing countries [30].

Electronic medical records (EMR) systems, with inclusion of patients through access to individual personal health records not only improve quality of health care given to individuals and reduce cost of health care in a country, but also engage individuals in their own health care [31], both in private practice as well as their adaption at a national level [32]. Benefits of computerized Information Systems for the information needs of the healthcare system are countless, particularly in developing countries [33]. Implementing e-health in developing countries showed significant improvement in ability to track patients, monitor adherence of patients to the treatment regime, and keep track of those who do not follow up their treatments and appointments [19]. The time consumed in communication significantly drops due to use of e-Health. It is important to note the difference between e-Health and EMR. E-Health is the transfer of health resources and health care by electronic means [34] whereas an EMR contains the standard medical and clinical data gathered in one provider's office [35]. Essentially, e-Health has the capability to extend the uses of an EMR system by sharing it electronically. EMR systems help developing countries by effectively facilitating data collection, data entry, information retrieval, report generation and research [33]. Real-time application of EMR contributes towards effective clinical decision support [36], process automation with the potential to improve the quality of patient care and significantly reduce costs [37]. The use of touch screen systems and hand held devices such as personal digital assistants integrated with EMR systems are quite helpful in a developing nation like Malawi both in emergency medicine and field work such as collecting data for initiatives in proactive public health programs for example Human Immunodeficiency Virus Awareness (HIV) programs [38]. The use of mobile applications and telecare in hospitals improves health status of patients, quality of care, reduces cost, and overall increased satisfaction of patients by allowing them to have access to knowledge about their health and their overall disease management [39] [40].

Fiji's current population stands at 837,271 [41]. There are a total of 186 government hospitals and health centers in Fiji that are staffed with a total of 550 government general practitioners [42]. However, according to Fiji Medical Practitioners Association, the current Patient to Medical Practitioner rate is at an alarming 1:1522. With the extreme shortage of medical professionals in Fiji [43] the intervention of the appropriate type of information technology becomes very essential.

This paper presents a case study of the current status of electronic medical records and health information systems in Fijian hospitals and health centers. The study aims to find if the general population of Fiji who has access to Internet would prefer to have access to electronic record systems. We review the current EMR systems in Fijian hospitals and study its strengths and limitations. Nausori Health Center is used as a case study for this paper. It also attempts to find out if the medical practitioners of Fiji as well as the general public would be interested in a cloud based EMR system with a patient portal for online appointments and reports. Surveys were used to understand the current state of Health Informatics in Fiji, and to establish if Fiji has enough technology and resources to introduce a free and open source EMR system that can be accessed in any government hospital or though smart phones.

# Methods

This study uses grounded theory with an exploratory approach [44] [45] [46]. It started by trying to understand the perceptions of the users towards the information management practices in Fijian hospitals, and to learn if they are currently using any EMR. Qualitative study was conducted by using two primary methods of research, which are, surveys [10] and interviews [47] to understand user perspectives about information systems in Fijian hospitals. These surveys had questions pertaining to their experiences when they visit hospitals (patient/public survey) and the experiences of the staff of government hospitals when dealing with information within their facilities.

The surveys were conducted both online (Sogo Survey) and through hard copies while interviews were done during visits to the health facilities including Ministry of Health, Fiji. All participants interviewed or surveyed remain anonymous as the research team had signed agreements for confidentiality and ethics for working with the staff of Ministry of Health. The conclusions were derived after analysis of the results in the data tables through Microsoft Excel files that were generated by Sogo Survey.

The original grounded theory approach was used since the study has potential to reveal more results than what was originally anticipated [44].

**i) Surveys**

There were two different type of surveys conducted:

a) ***Patient/General Public Survey-*** the data collection sheet was given to the general public and patients both in hard copy and as a link to an online version of the survey. 216 people had participated (n= 216 as these were the maximum number that consented to participate in this survey)

b) i) ***Staff Survey-*** for Nausori Health Centre- this survey was given only to the workers of Nausori Health Centre (n= 23 as this was the total number of staff that had consented to participate)

ii) ***Staff Survey-*** for other government health centres/hospitals throughout Fiji- this survey was exactly like the survey for Nausori Health Centre; however this was done to compare the results of Nausori Health Centre to the feedback of staffs with rest of the government hospitals in Fiji. (n= 100 as these were the maximum number that consented to participate in this survey) The participants were from the main hospitals in Suva, Nadi, Lautoka, Labasa areas of Fiji.

**ii) Interviews**

Interviews were carried out with the ICT team of Ministry of Health and UNAIDs[1] information advisors in order to further understand the current information systems used in the health facilities in Fiji. Nausori Health Centre was a case study for this research as it serves up to 300 patients per day. It is also the largest health centre in the Rewa Subdivision in Fiji.

# Results

## Results from Assessment

### i) Case-Study: Nausori Health Centre

Nausori Health Center was chosen for this case study since it serves a large population, which includes two provinces of rural communities, therefore this study helped gauge the information systems needs of an average government hospital in Fiji. Nausori Health Center serves as the major government hospital, in Rewa Subdivision in the Central area of Viti Levu, Fiji. It comprises of various departments such as dental, maternity, public health, general outpatient for accident and emergency, special outpatients for chronic diseases, integrated management of childhood illness, triage, pharmacy and administration. Every patient that is served at Nausori Health Center has a unique National Health Number (NHN) which is generated by a system called Patient Information System (PATIS). This number is the national ID that is used in all hospitals. This system is used only to generate a NHN for a new patient, or to check the NHN of an existing patient. If the patient is an existing patient, then a search is done to know the NHN. Once the NHN is found, the staff manually locates a corresponding manila folder which has manual records of the patient's medical history, prescriptions, history of previous visits and all other information pertaining to the patient. In case it is the patient's first visit, PATIS is used to generate a new NHN, which is then printed out and physically pasted with adhesive paper glue on a new manila folder to follow the same practice as that of the existing patients. There is a huge room with all the manila folders that are labelled according to the initial of the patient's last name. It could take anywhere between 10 minutes to an hour to successfully locate the patient's folder. Even in that case, it is highly unlikely that the folder may contain all accurate information about the patient, as most of the medical practitioners do not have time to do proper data entry, manually or electronically. In case a patient has to visit any other government medical center or hospital in Fiji, they have to take their folders themselves.

[1] UNAIDS-Joint Nations Programmed on HIV/AIDs

Such a practice is allowing room for anomalies at many levels. The entire practice is clearly too time consuming and cumbersome for both the patients and the staff of the health center. Not having access to timely, accurate and consistent data, delays diagnosis and treatment procedures for the patients. Due to such practice, the records are not accurate or up to date. According to the responses collected from interviews, the practice of patient's having access to their own folders may raise concerns of security and data manipulation. Relying on an almost manual system makes it very difficult to deal with complains as records are not entered well, therefore investigations can go on for up to a year. This was evident in a case in one of the hospitals where a patient is suing Ministry of Health for negligence [48]. The case is still under investigation for over one year. Having an EMR system would provide some evidence for the case to be solved in far less time compared to how complaints are dealt with now. It is clear that Heath Information infrastructure in Fiji currently remains inadequate to meet the needs of Fiji's population.

PATIS, the current Health Information Systems used by Government hospitals in Fiji, is available mainly in the central/eastern, western and northern parts of the country. It is yet to be implemented in all government health centers across Fiji. The PATIS system was adapted based on a HIS in Samoa. It was jointly funded by the Australian and the Fijian Government in 2001.

### ii) HIS and PATIS in Fiji

According to the interviews conducted with Ministry of Health, there are a total of 200 government hospitals and health centers in Fiji. Out of this, only 36 centers are linked with a health information system called Patient Information Systems (PATIS). The remaining centers still use paper based manual systems. PATIS is the primary software used as an EMR. The Ministry of Health is currently in the process of upgrading to PATIS Plus, which is an improved version of the earlier PATIS software. PATIS plus is intranet based, with the possibility of having remote access via Virtual Private Network tunnels. It has basic functionality of an EMR which includes patient administration, details on medication, admissions and discharge. Apart from PATIS/PATIS Plus, the hospitals are also using PHIS (Public Health Information System), a reporting tool used within the hospital for nursing stations (not at patient level), LabIMS for lab tests and RIS (Radiology Information System) for x-ray systems. Inventory and finance is managed by software called Epicor. All these software have been provided by AusAID [42]. Even though there are 36 live sites equipped with PATIS, it has been found that most of the centers do not use any software at all, as they are understaffed and have very few computers to work with.

**Modules of PATIS Plus**

These are the modules of PATIS Plus system:

- Patient Master Index- PMI- contains basic patient demographics and allows to search and view patients records. This is the first interface of the system
- General Outpatients Department - GOPD- shows information about the processes from triage to diagnosis for general outpatients

- Special Outpatient Department - SOPD- shows information about specialty cases such as accident and emergency, special clinics diabetics, eye clinic and family clinic
- Admission and Treatments Department- ATD- shows information about admission for patients
- Pregnancy and Birth- P&B- shows information about ante natal to full birth cycle
- Surgery- shows all data pertaining to surgical procedures
- Pharmacy- shows information about medication inventory and dispensing
- Radiology- holds data on bookings for radiology (linked RIS)
- Appointment- this module is used to decide about appointments and to make bookings
- Reports (PATIS Plus reports of any module) and External Systems (Lab IS, RIS, other software are interoperable with PATIS plus)
- Site Administration- creates and manages users, resets passwords

There are a total of 200 health centers and hospitals in Fiji. Out of that, 36 sites are now live with the new PATIS Plus.

Other stations with PATIS Plus include:
1. Ministry of Health - Head Quarters
2. Diabetes Hub Center – Central Division
3. Western Health Services Office
4. Northern Health Services Office

## Results from Surveys

### Feedback from Patients and Government Health Workers

This section analyses the current state of EMRs in Fiji. The study was done by conducting surveys on
(i) patients and general public, with 216 participants,
(ii) staff of Nausori Health Center, which is the case study, with 23 participants
(iii) Staff of general health centers and hospitals in Fiji, with 100 participants from major hospitals and health centers in Suva, Nasinu, Lautoka, Nadi and Labasa.

The sole purpose of the survey was to compare results of Nausori Health Center to the rest of the health workers in Fiji.
The surveys were launched online using Sogo Survey[2]. The link to the online survey was distributed through emails and social networking sites and was also available in printed copies for those who did not have access to Internet. All participants of the survey were anonymous.

Below are the results from all 3 surveys. The results of the surveys are discussed below:

### i) Survey for General Public and Patients of Government Health Centers and Hospitals

This section shows results from the survey that was conducted for general public and patients of government hospitals in Fiji. They were surveyed on their experience while visiting government hospitals in terms of information systems, if they are smartphone/internet literate, and if they would like to make online bookings for their appointments.

[2] https://www.sogosurvey.com/k/SsQWYWRsVsPsPsP

***Survey Questions*** ***Survey Participants = 216***

| |
|---|
| Q1. Do you own a smartphone? |
| Q2. Do you have access to Internet/email (including social networking sites like Facebook)? |
| Q3. Do you use a nationally recognized government medical card (called National Health Card)? |
| Q4. How long does it take the staff to find your information (medical records) with the card? |
| Q5. How long does it take the hospital staff to find your information (medical records) without the card? |
| Q6. Would you like to make your appointments for check-up electronically (through Internet or smart phone)? |
| Q7. When transferred from your local government hospital to another government hospital (eg. Nausori Health Centre to Colonial War Memorial Hospital), did the second government hospital had your medical records prior to your visit |
| Q8. When transferred from your local government hospital to another government hospital, how was your information available prior to your visit? |
| Q9. Many countries in the world use a web based, free and open source health information system software, which can be accessed by medical practitioners(complete Electronic Medical Records system)as well as patients (to view their reports and make appointments) throughout the country. Would you like such a system in the government hospitals in Fiji? |

*Table 1 Survey Questions for General Public & Patients of Government Health Centres & Hospitals*

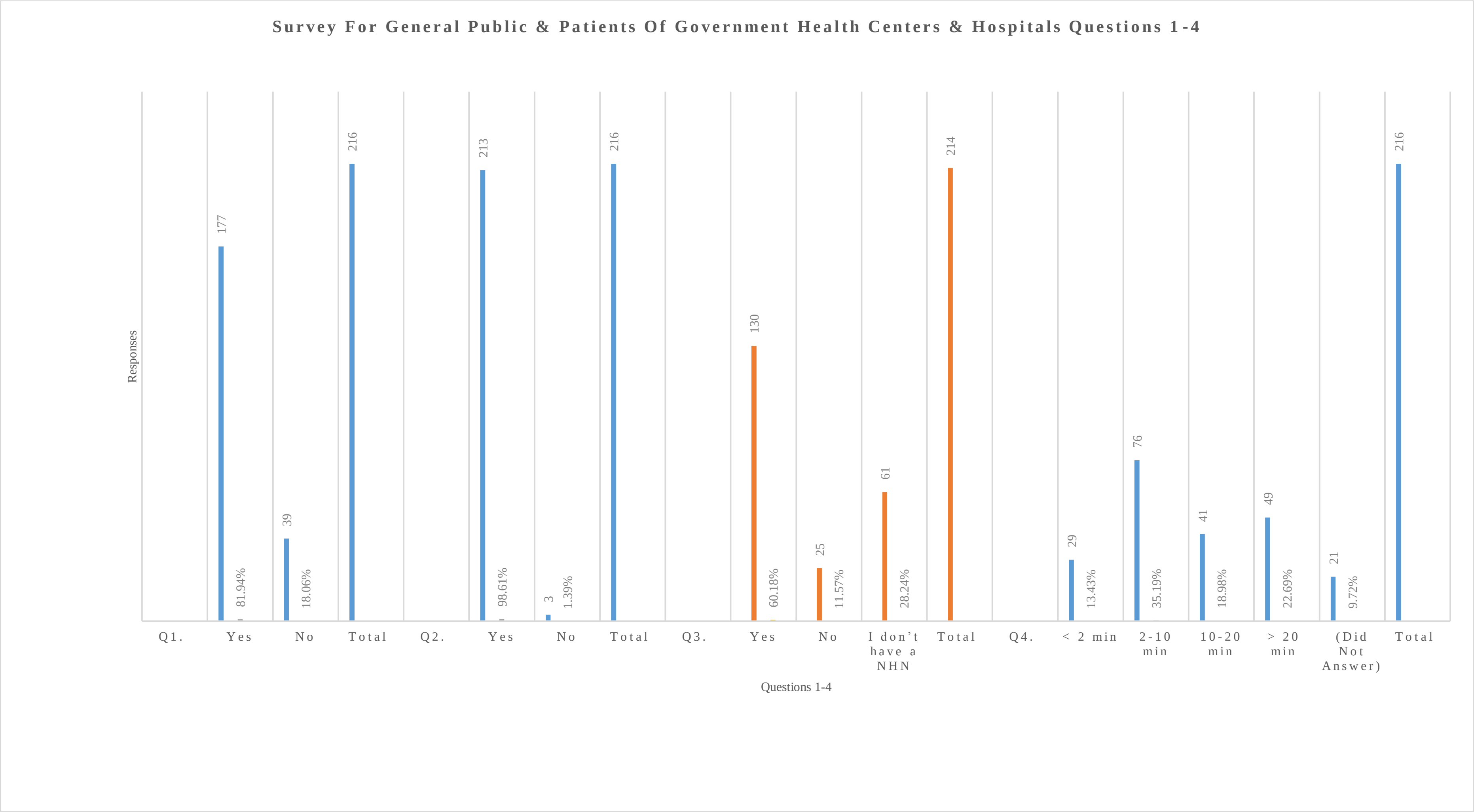

*Figure 7 Results of the Survey for General Public and Patients of Government Health Centres and Hospitals Questions 1-4*

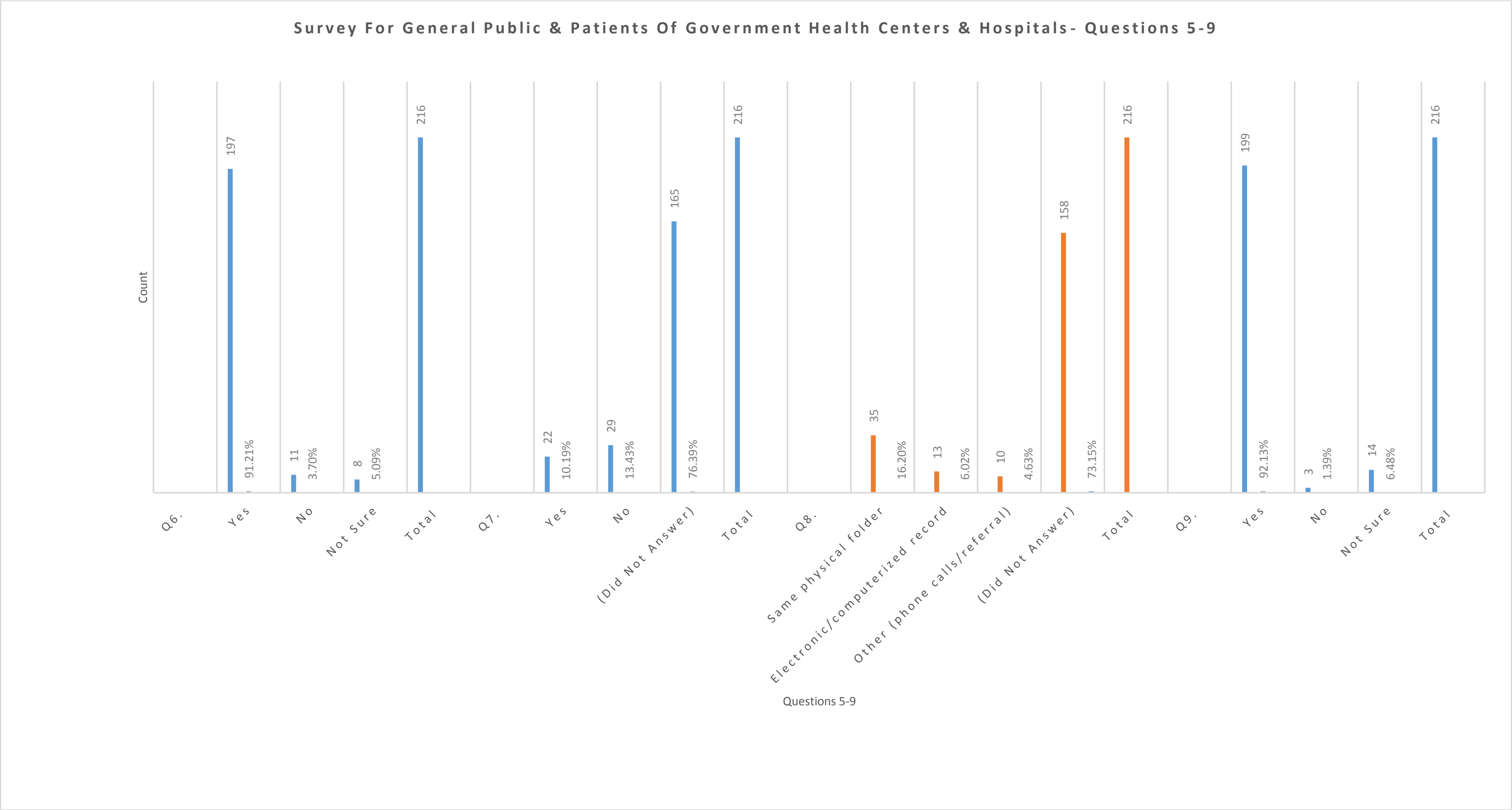

*Figure 8 Results of the Survey for General Public and Patients of Government Health Centres and Hospitals Questions 5-9*

According to the results in Figure 8, 81.97% of the survey participants have smart-phones. 98.61% of the participants have access to Internet/email/social networking sites. 60.18% have National Health Cards. It is alarming to note that 35.19% patients observed that it takes the administration staff between 2-10 minutes to find their records if they have their National Health Cards. Howsoever, 46.76% patients have responded that it takes more than 20 minutes if they do not have the National Health Cards with them.

Another concern is that 10.19% of patients said that while being transferred from to another hospital, the second hospital had their information prior to the transfer. However only 6.02% said that the second hospital had their information in electronic medical records. The survey results also show that 91.21% of the participants would like to make their appointments and access reports online (patient portal) or via their smartphones, while 92.13% of the participants are favourable of the idea of having a free and open source EMR system which can be accessed anywhere in Fiji, as a universal solution for Health Informatics in Fiji

### ii) Survey for Staff of Government Hospitals

The following section shows results from the survey that was conducted at Nausori Health Center. They were surveyed on their current practices in dealing with information within their hospital such as patient information and internal information such as patient medical records and transfer information, reporting smartphone/internet literacy, and if they would like to make bookings for their appointments online.

#### a) Responses from Nausori Health Center (Case Study)

***Survey Questions*** ***Survey Participants = 23***

| |
|---|
| Q1. Do you own a smartphone? |
| Q2. Do you have access to internet/email (including social networking sites like Facebook)? |
| Q3. Please explain why you do not use computers to do your work |
| Q4. Please describe how the information is entered/stored |
| Q5. What mechanisms do you currently use in order to complete your reports and disseminate information? |
| Q6. Do you think tablet computers will help you work more efficiently? |
| Q7. Would you like to be able to do electronic online/web based booking from any physical location in Fiji, for the patients you attend to? |
| Q8. While transferring a patient from your hospital to another hospital did the second government hospital have all information of the patient? How was the information available? |
| Q9. Many countries in the world use a web based, free and open source health information system software, which can be accessed by medical practitioners (complete Electronic Medical Records system) as well as patients (to view their reports and make appointments) throughout the county. Would you like such a system to be used in Fiji? |
| Q10. Do you have access to the PATIS Plus system |
| Q11. What are the problems you face with using PATIS Plus? |
| Q12. Many countries have integrated electronic medical record system that allows patient records to be shared between the practitioners of government hospitals as well as private practitioners. Such a practice allows faster, convenient access to patient records and more efficient reporting. Do you think such an integrated system will be good for Fiji? |

*Table 2 Survey Questions for Staff of Nausori Health Centre (Case Study)*

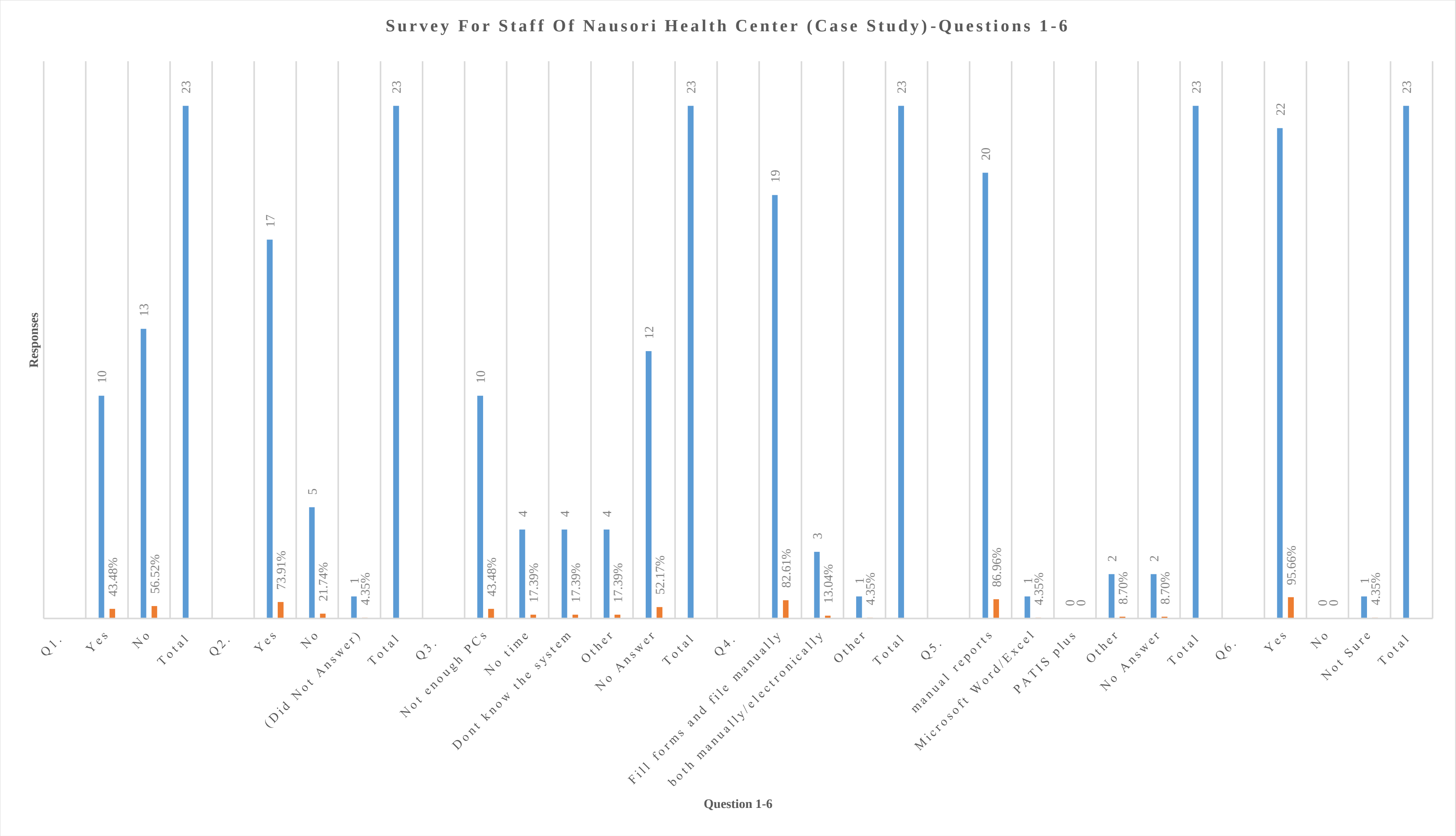

*Figure 9 Survey for Staff of Nausori Health Centre (Case Study)- Questions 1-6*

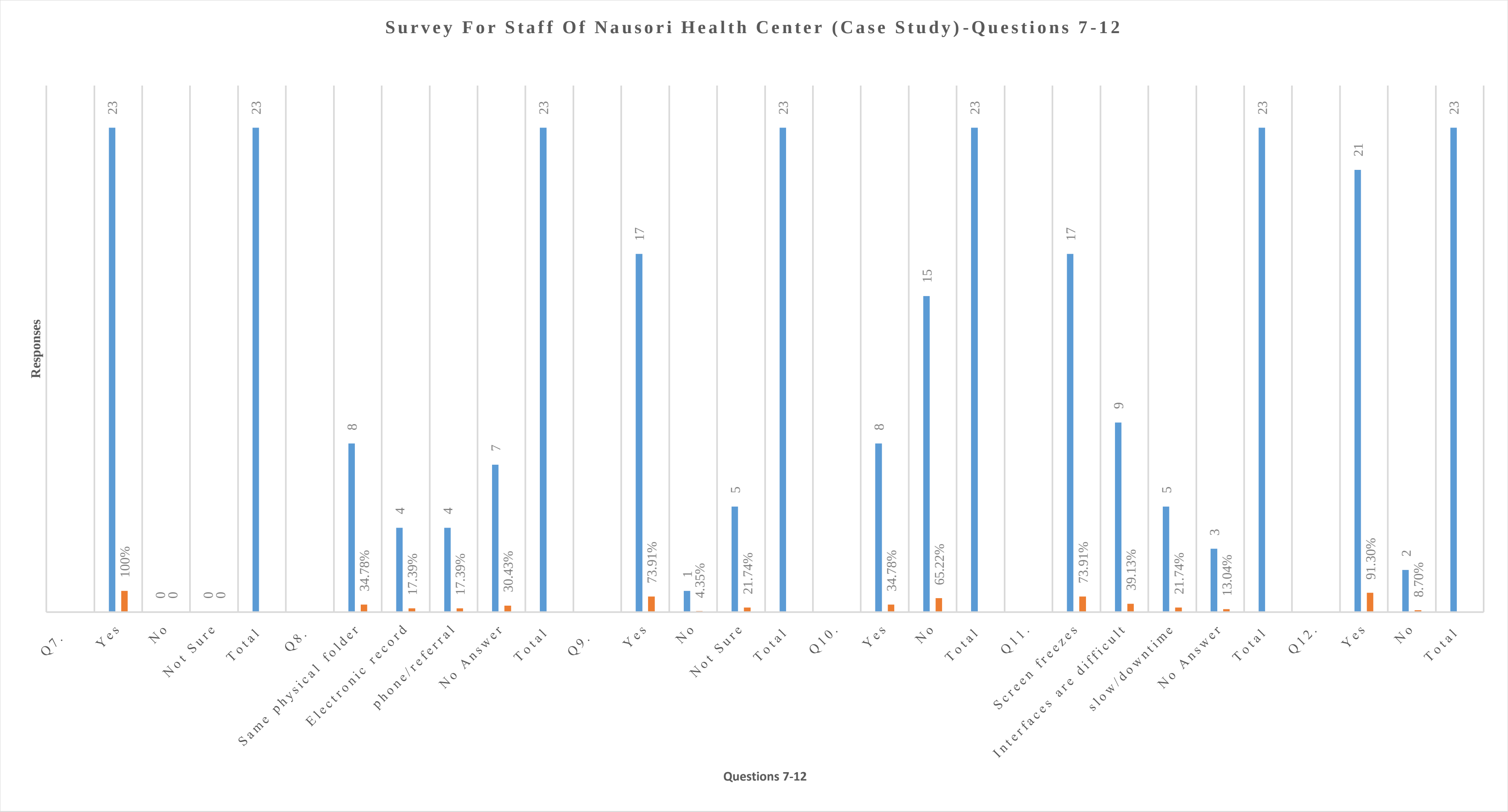

*Figure 10 Survey for Staff of Nausori Health Centre (Case Study) - Questions 7-12*

According to the results in Figure 9, 43.48% of the survey participants have smart-phones. 73.91% of the participants have access to Internet/email/social networking sites. 52.7% say they do not have access to computers. 82.61% say that information is currently entered or stored by filling in manual forms and filing them in manila folders, while 86.96% say that they create reports manually. 95.66% think that Tablet computers will help them work more efficiently. 17.39% say that when transferring a patient from one hospital, the second hospital had electronic medical records of the patient prior to the patients visit. 73.91% of the participants would like to make their appointments for their patients online or via smart-phones using an EMR system which can be accessed anywhere in Fiji, as a universal solution for Health Informatics in Fiji. Furthermore, 34.78% have access to PATIS Plus, however, 73.91% of them responded that the screen freezes frequently, which is a major limitation of the software. 91.30% responded that they would like to have an integrated EMR that allows patient records to be shared between private and government practitioners. Overall, the participants wanted changes such as more computers, trained staff, user friendly systems for the staff with special needs, 24/7 access to systems, timely data entry and record keeping for better decision making and to have information about patient medication.

**b) Responses from Staff of other Government Hospitals (Generally)**

The following section shows results from government hospitals throughout Fiji. Participants included staff of government hospitals and health centers from central, western and northern division. They were surveyed on their current practices in dealing with information within their hospital, such as patient medical records, patient transfers and reporting. They were also asked if they are smartphone/internet literate and if they would like to make bookings for their appointments online.

***Survey Questions*** ***Survey Participants = 23***

| |
|---|
| Q1. Do you own a smartphone? |
| Q2. Do you have access to internet/email (including social networking sites like Facebook)? |
| Q3. Please explain why you do not use computers to do your work |
| Q4. Please describe how the information is entered/stored |
| Q5. What mechanisms do you currently use in order to complete your reports and disseminate information? |
| Q6. Do you think tablet computers will help you work more efficiently? |
| Q7. Would you like to be able to do electronic online/web based booking from any physical location in Fiji, for the patients you attend to? |
| Q8. While transferring a patient from your hospital to another hospital did the second government hospital have all information of the patient? How was the information available? |
| Q9. Many countries in the world use a web based, free and open source health information system software, which can be accessed by medical practitioners (complete Electronic Medical Records system) as well as patients (to view their reports and make appointments) throughout the county. Would you like such a system to be used in Fiji? |
| Q10. Do you have access to the PATIS Plus system |
| Q11. What are the problems you face with using PATIS Plus? |
| Q12. Many countries have integrated electronic medical record system that allows patient records to be shared between the practitioners of government hospitals as well as private practitioners. Such a practice allows faster, convenient access to patient records and more efficient reporting. Do you think such an integrated system will be good for Fiji? |

*Table 3 Survey Questions for Staff of other Government Hospitals in Fiji*

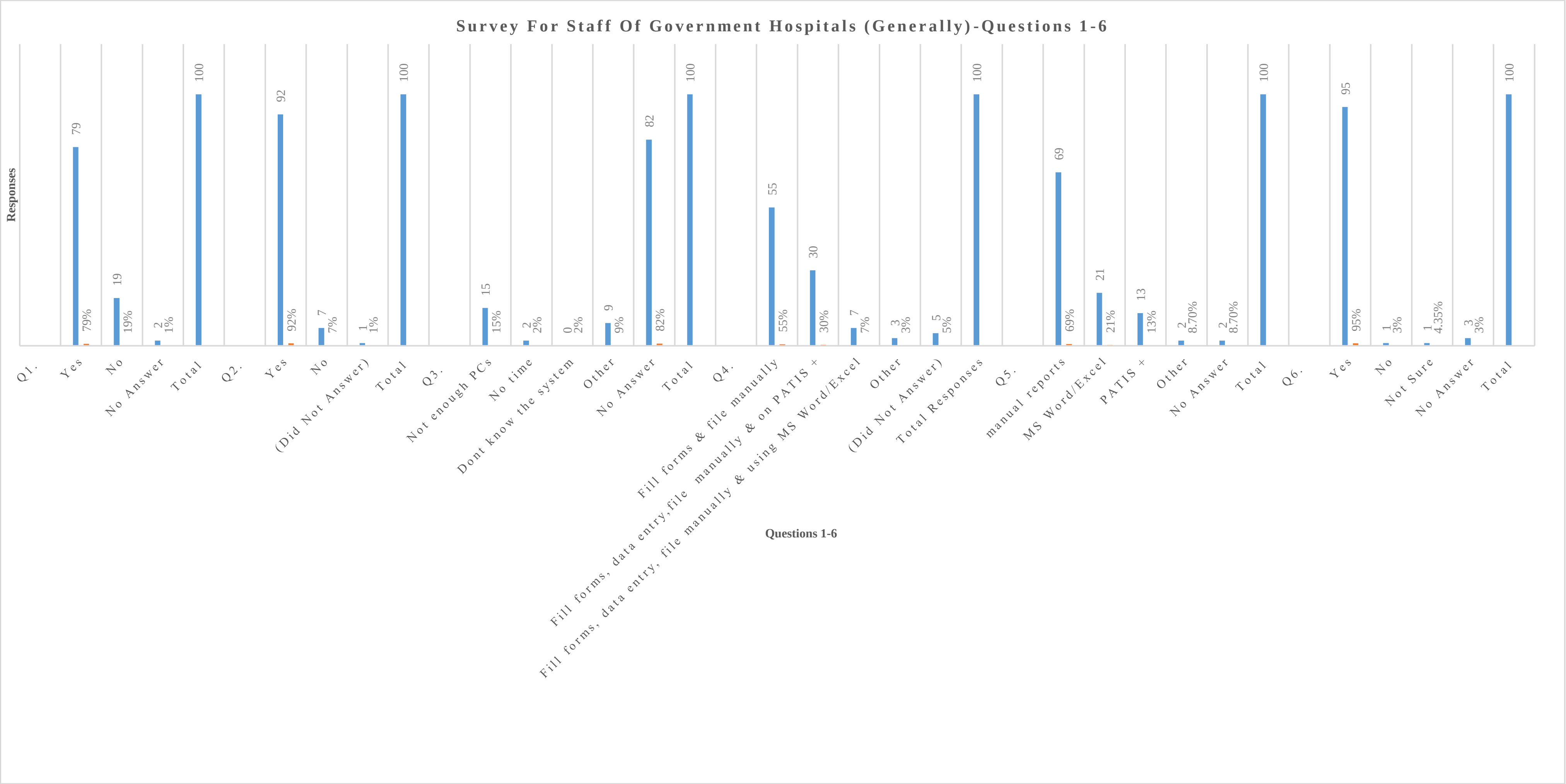

*Figure 21 Responses from Survey for Staff of Government Hospitals and Health Centres (Generally) Questions 1-6*

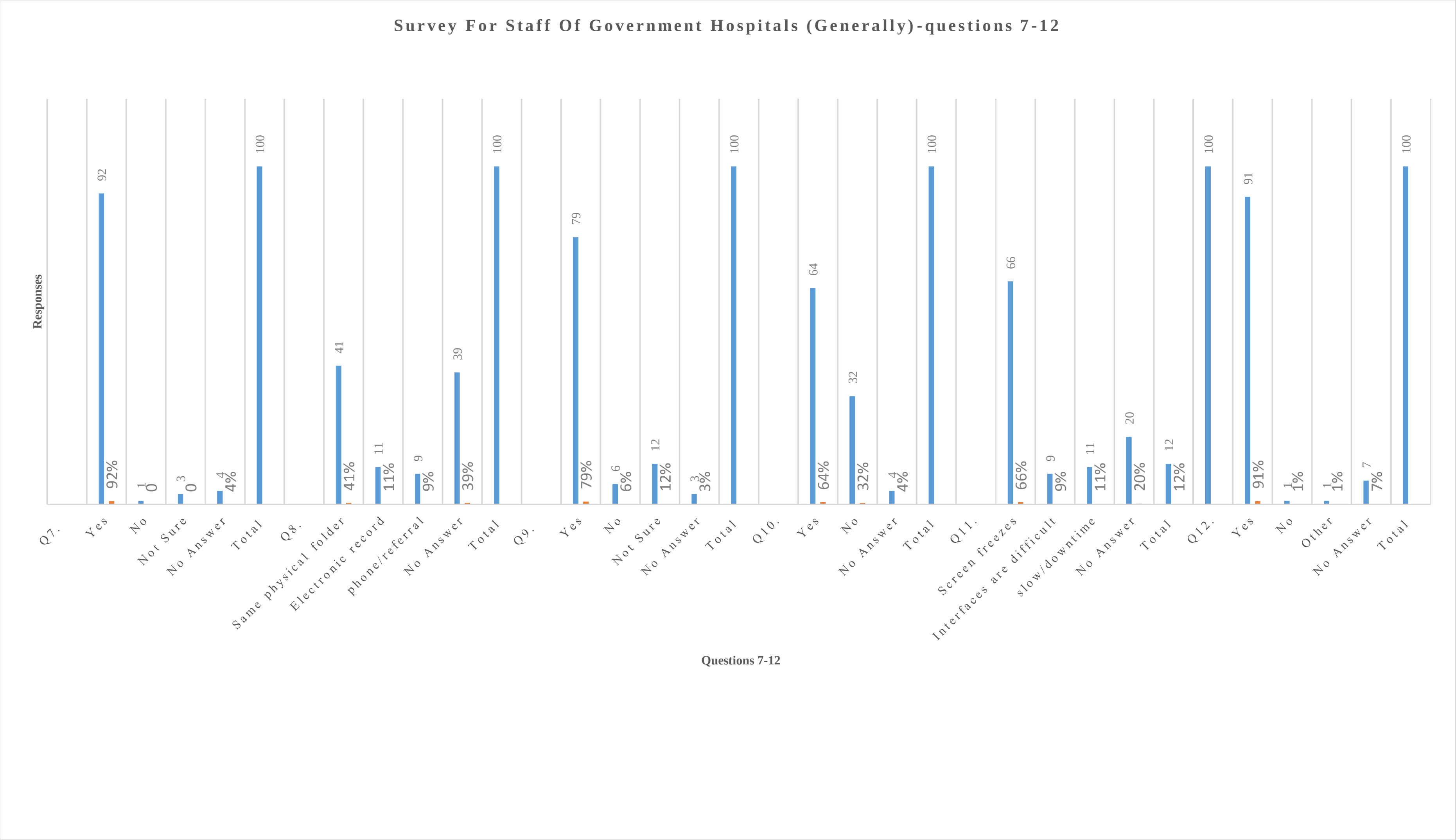

*Figure 3 Responses from Survey for Staff of Government Hospitals and Health Centres (Generally) Questions 7-12*

According to the results above, 79% of the survey participants have smart-phones. 82% of the participants have access to Internet/email/social networking sites. 55% say they do not have access to computers. 69% of the hospital staff say that information is currently entered or stored by filling in manual forms and filing them in manila folders, while 95% responded that they create reports manually. 92% agree that Tablet computers will help them work more efficiently while a minor 11% say that when transferring a patient from one hospital to another, the second hospital had electronic medical records of the patient prior to the patients visit. 79% of the participants would like to make their appointments for their patients online or via smart-phones using an EMR system which can be accessed anywhere in Fiji, as a universal solution for Health Informatics in Fiji. 64% of those who participated in the survey have access to PATIS Plus however 66% say that the screen freezes frequently. A majority of 91% say that they would like to have an integrated EMR that allows patient records to be shared between private and government practitioners.

Overall, the participants wanted changes such as more computers, trained staff, user friendly systems, no downtime, better stock management, 24/7 access to systems, timely data entry and record keeping for better decision making and visibility into information about the medications. The results of all the surveys indicated that majority of the participants (patients, general public, staff of Nausori Health Center as well as staff of other hospitals around Fiji) want a Health Information Systems such as an Electronic Medical Records system to manage patient data at government hospitals in Fiji. Patients as well as medical practitioners were favourable of the idea of having access to health records available on a web based Electronic Medical Records system which can be accessed by any government hospital in Fiji. They also wished to make their appointments for check-up online using smart phones and mobile applications

**Part 2 Feedback from Interviews[3]**

All interviews conducted were confidential, therefore no names are mentioned. Majority of the participants wanted an Electronic Medical Records system to manage patient data in government hospitals in Fiji. Patients and medical practitioners were favourable of the idea of having access to web based Electronic Medical Records system. They also want to make online appointments for check-up. It is evident that most health facilities in Fiji are under resourced in terms of computers. In some cases there are only 2 or 3 computers in the entire health center. Most health centers have only 2 computers, used at reception and pharmacy respectively. The same is the case of Nausori Health Center as well. The medical practitioners have mentioned that they face a lot of difficulties in addressing the needs of their internal customers and stakeholders. For example, they only learn of a particular medication being out of stock when they check manually. Thereafter, it takes months to procure the much needed medication from suppliers. Similarly, other areas of the hospital management and administration are in dire need of a fully integrated Information System Solution to automate most of their manual tasks and to operate in an efficient and effective manner. Furthermore, the lack of finances has long impeded the initiatives of exploring quality Health/Hospital Information Systems. Such solutions can be expensive, particularly for developing countries like Fiji. Overall, the results of the surveys and interviews indicate that the people of Fiji have enough access to technology to be able to use an online Electronic Medical Record System, such as OpenEMR. However, some participants have raised concerns about the security of information, if free and open source software was to be implemented. They have also asked what level of access will the public have to the system. An overwhelming finding of the survey indicates that government health centers across the nation lack computers facilities. There is limited access to PATIS Plus and there is only a handful of staff trained to use PATIS Plus. Existing PATIS Plus users also find it difficult to use the application as it is generally slow due to a low bandwidth. PATIS Plus is still in its implementation phase [49]. Not all sites are live yet, as they would require additional resources such as access to computers and internet. According to Ministry of Health, currently, 36 out of 200 sites have PATIS Plus deployed; however it is not being fully utilized yet. Interviews with Ministry of Health state that there are no limitations in the PATIS Plus software, however, any ICT based tool may not solve the issues of record keeping in government hospitals. Processes need to be designed and mapped using the tools in order for a health information to be successful. PATIS plus is reviewed quarterly and the information is consolidated into a report, which is acted upon in order to rectify any issues that have been reported. PATIS Plus is funded by AusAID, however, hardware and support is not funded [42]. Therefore, it is important to invest in the procurement of hardware, training and technical support in order to realize the full potential of PATIS Plus.

[3] Under the Ethics/Confidentiality agreement, participants of all interview are anonymous

# Discussion

### i) Health Information Systems in Fiji

The lack of computers makes it difficult for most staff to do data entry or check records of the patient's medical history; therefore, they continue to use manual systems. The majority of the staff surveyed and interviewed have expressed a need for having computers for better records keeping and information management. There is a general shortage of qualified medical practitioners in Fiji [49]. The lack of a fully computerized Health Information Systems is adding to the woes of the already overloaded staff in the government hospitals in Fiji. Manual systems lead to cumbersome paperwork and long delays in attending to the patients. The field of Health Informatics has become a significant field of study due to the various challenges faced by the medical field. Internationally, there has been a rapid growth in the volume of medical knowledge and patient information, due to advancement in more treatments and interventions. These have coincidently produced more information management needs. However, there is a need for the ability to catalogue the information to be archived for future purposes. In addition to this, information is not standardized, which creates provisions for further anomalies when accessing information about the same patient from different hospitals around the country.

There is obviously an urgent need for a highly versatile and efficient Health/Hospital Information System, which will be an enterprise level solution to cater for the various information management needs of any hospital.

### ii) PATIS Plus

Not many health workers have access to PATIS Plus or a computer; however, if needed, they are able to access PATIS Plus from the computers in administration department, with the login details of the PATIS Plus operator. Generally health workers are finding PATIS Plus slow, however many will be satisfied if they have their own computer with their own access, 24/7. Many health care centres are still waiting for PATIS Plus to be implemented so it can be used by all health workers across the country.

### iii) Web based EMR

Most of the participants of the surveys are favourable to the idea of having an online web based EMR system that can be accessed by practitioners anywhere in Fiji. Participants would also like an integration of the EMR between private and government participants over a cloud based system. However, some have raised security concerns if some integration of information were to happen.

Large number of the participants show interest in a web based EMR system for hospitals in Fiji. Participants have shown interest in having a patient portal for reports and booking. They have also given favourable responses to the idea of sharing their medical records between private and government hospitals for ease of reporting resulting in faster services as well as online access for health workers to allow mobility. However, as indicated in results, a major limitation of the Information Systems in Fijian hospitals is the lack of computers in general. Further most medical practitioners do not have time to do data entry. However they do think applications on smart phones could help them to enter data entry easily and conveniently.

This study can guide medical practitioners and stakeholders in gaining a better understanding of HIS, which can help them in establishing feasible and appropriate solutions for managing information in public hospitals. This can potentially result in making better, informed decisions relating to disease treatment and prediction of the trends in outbreak of epidemic diseases. This research will pave a pathway for all Information and Communication Technology and Information Systems professionals for any future Health Informatics project.

# Conclusions

In this paper, we presented an assessment of the status of health informatics in Fiji through interviews and surveys conducted in Fiji with the patients, medical practitioners and staff of Ministry of Health. Our results show that a fully integrated web based EMR system is needed to fulfil the computational and technological requirements of public hospitals in Fiji. The obstacle in place has been the lack of computers in Fijian government hospitals. Additional security measures may need to be implemented for web based EMRs. Patient confidentiality, organizational ICT policies and the country's legislation need to have such concerns addressed at their respective levels to support the technical measures. The case study on Nausori Health Center reveals that there are still many manual practices that is causing unnecessary delays for both patients and workers. The findings suggest that the majority of the staff at Nausori Health Center would appreciate a complete online EMR system to handle patient records and reporting. This study has established that government health workers are interested in a web based EMR system that has a patient portal, so patients can access their reports and make appointments for their check-up themselves. However, further study is necessary to understand web based EMR systems that would meet the requirements of a developing nation such as Fiji. EMR systems should be made accessible through smart phones and Tablets even in remote locations such as the outer Fiji islands.

### Findings

Nausori Health Centre is still operating manually. The only area that consistently uses the PATIS Plus system is the pharmacy department. Most patients are served using manual records. A separate book is kept for patients who visit the hospital at nights (this includes emergencies), after the PATIS Clerk finishes her duty. Most of the hospitals and health centres in Fiji face the same predicaments; however it seems the largest hospital, Colonial War Memorial Hospital, uses PATIS Plus extensively compared to all the hospitals and health centre in Fiji. The users may increase if more computers are provided and infrastructure is made available to all centres.

Major findings of this study were:

1. Most health facilities in Fiji operated manually in terms of information systems

2. There is a general shortage of computers in government health facilities in Fiji

3. PATIS Plus was generally slow and very few users had access to PATIS Plus. Many users shared the same login as that of the PATIS Plus operator due to shortage of computers and individual accounts

4. The staff of government hospitals in Fiji think that they will be able to operate more efficiently if they had their own computers with their own logins for PATIS Plus

5. There are inconsistent practices in terms of information systems management across government hospitals in Fiji

6. Even with full implementation of PATIS Plus in all centres, the software may not necessarily be successful as it may need some business process re-engineering in order to meet the needs of Fijian hospitals

7. The people of Fiji (health practitioners and general public) have enough computer and internet access and literacy and are ready for a web based electronic medical records system that can be accessed through smart phones in any location. Patients would like to access their medical reports through their own portal which will also allow making online appointments.

### Limitations

It was difficult to interview the staff of government hospitals or ask them to take part in the surveys, as most health facilities in Fiji are under staffed and always busy. The results of the study would have been different had more participants volunteered to take part in the study. Moreover, after this study, it becomes imperative to conduct further user acceptability tests or black box testing on PATIS Plus to ascertain problem areas and suggest changes that could result in a higher user acceptance among staff of government hospitals.

### Future Work

This study can be extended to investigate the use of EMR systems through mobile applications in order to fully utilize the benefits of cloud computing in health sector. The use of Open Source EMR systems and prospectus for their implementation in developing nations can also be carried out. A collaborative study with Internet Service Providers in Fiji may be helpful in order to assess options for cloud infrastructure in Fiji.

## Authors' contributions

Main Author: Rohitash Chandra2* & Virallikattur S. Dhenesh1*

1 School of Computing, Information and Mathematical Sciences, University of the South Pacific, Laucala Campus, Fiji

2 Artificial Intelligence and Cybernetics Research Group, Software Foundation, Nausori, Fiji *

Authors are in order of contribution.

Email addresses:

SSR: swaran_ravindra@yahoo.com

RC: c.rohitash@gmail.com

VSD: vsdhenesh@gmail.com

# Acknowledgements

Ms Gabriela Ionascu, Strategic Information Adviser, UNAIDS

The Staff of Ministry of Health, Fiji

# References

[1] A. News. (2013, 21-April-2015). *Australian government criticised over AusAID merger* [News Article]. Available: http://www.abc.net.au/news/2013-09-19/an-ausaid-merger-reaction/4966976

[2] A. K. Y. a. N. J. B. R. E. Hoyt, *Health Informatics*, 5 ed. Florida: Lulu.com, 2012.

[3] P. M. Mell and T. Grance, "SP 800-145. The NIST Definition of Cloud Computing," National Institute of Standards and Technology2011.

[4] G. Eysenbach, "What is e-health?," *J Med Internet Res*, vol. 3, p. e20, 2001.

[5] HealthIT.gov. (2014, 02-January-2015). *What Is an Electronic Medical Record* Available: http://www.healthit.gov/providers-professionals/electronic-medical-records-emr

[6] T. EnterprisersProject.com. (2015). *What is Open Source*. Available: http://opensource.com/resources/what-open-source

[7] SourceForge. (2015, 10 July 2015). *SourceForge*. Available: http://sourceforge.net/projects/openemr/

[8] W. H. Organization. (2016, 25-June-2016). *WHO-About Us*. Available: http://www.who.int/about/en/

[9] M. B. Buntin*, et al.*, "The benefits of health information technology: a review of the recent literature shows predominantly positive results," *Health Aff (Millwood),* vol. 30, pp. 464-471, Mar 2011.

[10] R. Hillestad*, et al.*, "Can electronic medical record systems transform health care? Potential health benefits, savings, and costs," *Health Aff (Millwood),* vol. 24, pp. 1103-1117, Sep-Oct 2005.

[11] N. Menachemi and T. H. Collum, "Benefits and drawbacks of electronic health record systems," *Risk Manag Healthc Policy,* vol. 4, pp. 47-55, 2011.

[12] C.-F. Liu and T.-J. Cheng, "Exploring critical factors influencing physicians' acceptance of mobile electronic medical records based on the dual-factor model: a validation in Taiwan," *BMC medical informatics and decision making,* vol. 15, p. 4, 2015.

[13] A. Meißner and W. Schnepp, "Staff experiences within the implementation of computer-based nursing records in residential aged care facilities: a systematic review and synstudyof qualitative research," *BMC medical informatics and decision making,* vol. 14, p. 54, 2014.

[14] G. B. Cline and J. M. Luiz, "Information technology systems in public sector health facilities in developing countries: the case of South Africa," *BMC medical informatics and decision making,* vol. 13, p. 13, 2013.

[15] S. P. Sood*, et al.*, "Electronic medical records: a review comparing the challenges in developed and developing countries," in *Hawaii International Conference on System Sciences, Proceedings of the 41st Annual*, 2008, pp. 248-248.

[16] L. Raymond*, et al.*, "Improving performance in medical practices through the extended use of electronic medical record systems: a survey of Canadian family physicians," *BMC Med Inform Decis Mak,* vol. 15, p. 27, 2015.

[17] P. B. Jensen*, et al.*, "Mining electronic health records: towards better research applications and clinical care," *Nat Rev Genet,* vol. 13, pp. 395-405, 06//print 2012.

[18] J. P. Bansler and E. Havn, "Pilot implementation of health information systems: issues and challenges," *Int J Med Inform,* vol. 79, pp. 637-648, Sep 2010.

[19] J. A. Blaya*, et al.*, "E-health technologies show promise in developing countries," *Health Aff (Millwood),* vol. 29, pp. 244-251, Feb 2010.

[20] N. A. Mohammed-Rajput, *et al.*, "OpenMRS, a global medical records system collaborative: factors influencing successful implementation," *AMIA Annu Symp Proc,* vol. 2011, pp. 960-8, 2011.
[21] E. R. Weitzman, *et al.*, "Willingness to share personal health record data for care improvement and public health: a survey of experienced personal health record users," *BMC medical informatics and decision making,* vol. 12, p. 39, 2012.
[22] D. Gerdeman. (2014, 20-April-2015). *How Electronic Patient Records Can Slow Doctor Productivity*. Available: http://hbswk.hbs.edu/item/7452.html
[23] S. Y. Park, *et al.*, "The effects of EMR deployment on doctors' work practices: A qualitative study in the emergency department of a teaching hospital," *International Journal of Medical Informatics,* vol. 81, pp. 204-217.
[24] C. Mazzolini, "EHR holdouts: why some physicians refuse to plug in," *Med Econ,* vol. 90, pp. 42-4, 46, Nov 25 2013.
[25] J. D. Bramble, *et al.*, "The relationship between physician practice characteristics and physician adoption of electronic health records," *Health Care Manage Rev,* vol. 35, pp. 55-64, Jan-Mar 2010.
[26] J. M. McGrath, *et al.*, "The influence of electronic medical record usage on nonverbal communication in the medical interview," *Health Informatics J,* vol. 13, pp. 105-18, Jun 2007.
[27] J. Scholl, *et al.*, "A case study of an EMR system at a large hospital in India: challenges and strategies for successful adoption," *J Biomed Inform,* vol. 44, pp. 958-967, Dec 2011.
[28] F. Lau, *et al.*, "Impact of electronic medical record on physician practice in office settings: a systematic review," *BMC Med Inform Decis Mak,* vol. 12, p. 10, 2012.
[29] M. Khalifa, "Barriers to Health Information Systems and Electronic Medical Records Implementation. A Field Study of Saudi Arabian Hospitals," *Procedia Computer Science,* vol. 21, pp. 335-342, // 2013.
[30] A. Boonstra and M. Broekhuis, "Barriers to the acceptance of electronic medical records by physicians from systematic review to taxonomy and interventions," *BMC Health Serv Res,* vol. 10, p. 231, 2010.
[31] S. Mantwill, *et al.*, "EMPOWER-support of patient empowerment by an intelligent self-management pathway for patients: study protocol," *BMC medical informatics and decision making,* vol. 15, p. 18, 2015.
[32] P. C. Tang, *et al.*, "Personal Health Records: Definitions, Benefits, and Strategies for Overcoming Barriers to Adoption," *Journal of the American Medical Informatics Association : JAMIA,* vol. 13, pp. 121-126, 2006.
[33] F. Williams and S. A. Boren, "The role of the electronic medical record (EMR) in care delivery development in developing countries: a systematic review," *Inform Prim Care,* vol. 16, pp. 139-45, 2008.
[34] W. H. Organisation. (2015, 20-January-2015). *E-Health*. Available: http://www.who.int/trade/glossary/story021/en/
[35] P. Garrett and J. Seidman. (2011). *EMR vs EHR – What is the Difference?* Available: http://www.healthit.gov/buzz-blog/electronic-health-and-medical-records/emr-vs-ehr-difference/
[36] F. North, *et al.*, "Clinical decision support improves quality of telephone triage documentation-an analysis of triage documentation before and after computerized clinical decision support," *BMC medical informatics and decision making,* vol. 14, p. 20, 2014.
[37] B. Rothman, *et al.*, "Future of Electronic Health Records: Implications for Decision Support," *Mount Sinai Journal of Medicine: A Journal of Translational and Personalized Medicine,* vol. 79, pp. 757-768, 2012.
[38] C. f. D. C. a. Prevention. (2014, 02-January-2014). *Innovative Electronic Medical Record System Expands in Malawi*. Available: http://www.cdc.gov/globalaids/success-stories/innovativemalawi.html
[39] A. S. M. Mosa, *et al.*, "A systematic review of healthcare applications for smartphones," *BMC medical informatics and decision making,* vol. 12, p. 67, 2012.

[40] C. Free, *et al.*, "The Effectiveness of Mobile-Health Technology-Based Health Behaviour Change or Disease Management Interventions for Health Care Consumers: A Systematic Review," *PLoS Medicine*, vol. 10, p. e1001362, 2013.
[41] F. B. o. Statistics. (2015, 20-April-2015). *2007 Census of Population and Housing*. Available: http://www.statsfiji.gov.fj/index.php/2007-census-of-population
[42] F. N. U. Graham Roberts, *et al.*, "The Fiji Islands health system review " Asia Pacific Observatory on Health Systems and Policies ; World Health Organization, Manila : Geneva2011.
[43] M. o. H. Fiji, "Annual Report 2013," Ministry of Health2013.
[44] B. G. Glaser, *et al.*, "The discovery of grounded theory; strategies for qualitative research," *Nursing Research*, vol. 17, p. 364, 1968.
[45] A. Shachak and S. Reis, "The impact of electronic medical records on patient–doctor communication during consultation: a narrative literature review," *Journal of evaluation in clinical practice*, vol. 15, pp. 641-649, 2009.
[46] A. Bryant and K. Charmaz, *The Sage handbook of grounded theory*: Sage, 2007.
[47] A. L. Terry, *et al.*, "Perspectives on electronic medical record implementation after two years of use in primary health care practice," *J Am Board Fam Med*, vol. 25, pp. 522-527, Jul-Aug 2012.
[48] N. Swamy, "Back from the "Dead"," in *The Fiji Times*, ed. Suva, Fiji: The Fiji Times, 2014, p. 2.
[49] F. Ministry of Health, "PATISPlus Remediation Project," 1 ed. Suva, Fiji: Ministry of Health, Fiji, 2014, pp. 1-3.